\documentclass[preprint,aps,12pt,showpacs,nofootinbib,tightenlines]{revtex4}
\usepackage{amsmath}
\usepackage{amssymb}
\usepackage{CJK}
\usepackage{epsfig}
\usepackage{graphicx}
\usepackage{subfigure}
\usepackage{slashed}
\newcommand{\met}{\not\!\!\!E_{T}}
\begin{document}
\def\pslash{\rlap{\hspace{0.02cm}/}{p}}
\def\eslash{\rlap{\hspace{0.02cm}/}{e}}
\title {Double Higgs production in the littlest Higgs Model with T-parity at high energy $e^{+}e^{-}$ Colliders}
\author{Bingfang Yang$^{1,2}$}\email{yangbingfang@gmail.com}
\author{Guofa Mi$^{2}$}
\author{Ning Liu$^{1}$}
\affiliation{$^1$ College of Physics $\&$ Electronic Engineering,
Henan Normal University, Xinxiang 453007, China\\
$^2$ School of Materials Science and Engineering, Henan Polytechnic
University, Jiaozuo 454000, China
   \vspace*{1.5cm}  }

\begin{abstract}

In the framework of the littlest Higgs model with T-parity(LHT), we
investigate the double Higgs production processes
$e^{+}e^{-}\rightarrow ZHH$ and $e^{+}e^{-}\rightarrow
\nu\bar{\nu}HH$ at high energy $e^{+}e^{-}$ colliders. We calculate
the production cross sections and find that the relative correction
at the center-of-mass energy $\sqrt{s}=500$ GeV can maximally reach
$-30\%$ for the process $e^{+}e^{-}\rightarrow ZHH$ and $-16\%$ for
the process $e^{+}e^{-}\rightarrow \nu\bar{\nu}HH$ in the allowed
parameter space, respectively. These large relative corrections can
reach the detection range of the future $e^{+}e^{-}$ colliders so
that they can be used to test the LHT effect. The two relevant decay
modes $e^{+}e^{-}\rightarrow ZHH \rightarrow l\bar{l}
b\bar{b}b\bar{b}$ and $e^{+}e^{-}\rightarrow
\nu\bar{\nu}HH\rightarrow \nu\bar{\nu}b\bar{b}b\bar{b}$ are studied
and some distributions of the signal and background are displayed.

\end{abstract}
 \maketitle
\section{ Introduction}
\noindent

On the 4th of July 2012, ATLAS\cite{ATLAS} and CMS\cite{CMS}
collaborations have announced the existence of a Higgs-like
resonance around 125 GeV confirming the cornerstone of the Higgs
mechanism\cite{Higgs mechanism} that predicted such particle long
times ago. However, the discovery of a Higgs-like boson is not
enough to fully understand the mechanism of electro-weak symmetry
breaking (EWSB) and mass generation. The Higgs self-coupling is the
key ingredient of the Higgs potential and its measurement is
probably the most decisive test of the EWSB mechanism. To establish
the Higgs mechanism unique experimentally, the Higgs potential of
the Standard Model(SM)\cite{sm} must be reconstructed. In order to
accomplish this, not only the Yukawa couplings and the Higgs-gauge
couplings but also the Higgs self-couplings which include the
trilinear coupling and the quartic coupling should be measured.

The investigation of the Higgs self-couplings requires final states
containing two or more Higgs bosons. In fact, the cross sections for
three Higgs boson production processes are reduced by three order of
magnitude compared to those for the double Higgs boson production
\cite{H-pair1}\cite{H-pair2}, the quartic Higgs self-coupling
remains elusive. The phenomenology calculations show that it is
difficult to measure the trilinear Higgs self-coupling at the Large
Hadron Collider (LHC) due to the large QCD
background\cite{Hself-LHC}. But the $e^{+}e^{-}$ linear colliders,
such as the International Linear Collider (ILC)\cite{ILC} and the
Compact Linear Collider (CLIC)\cite{CLIC}, have clean environment
and can provide a possibly opportunity for studying the trilinear
Higgs self-coupling\cite{H-pair1}.

The littlest Higgs model with T-parity(LHT)\cite{LHT} was proposed
as a possible solution to the hierarchy problem and so far remains a
popular candidate of new physics. At the high energy $e^{+}e^{-}$
colliders, there are two main processes for the SM Higgs boson,
$e^{+}e^{-}\rightarrow ZHH$ and $e^{+}e^{-}\rightarrow
\nu\bar{\nu}HH$, where the former reaches its cross-section maximum
at a center-of-mass energy of around 500 GeV, while the
cross-section for the latter is dominating above 1 TeV and increases
towards higher energies. In the LHT model, some new particles are
predicted and some couplings of the Higgs boson are modified. These
new effects will alter the property of the SM Higgs boson and
influence various SM Higgs boson processes, where the double Higgs
production processes can provide a good opportunity to discriminate
between Product Group and Simple Group Little Higgs
models\cite{LH-DH}. The single Higgs production processes in the LHT
model have been investigated in our previous
work\cite{single-higgs}. In this work, we will study the double
Higgs production processes, $e^{+}e^{-}\rightarrow ZHH$ and
$e^{+}e^{-}\rightarrow \nu\bar{\nu}HH$.

The paper is organized as follows. In Sec.II we give a brief review
of the LHT model related to our work. In Sec.III we study the
effects of the LHT model in the double Higgs boson productions and
present some distributions of the signal and background. Finally, we
give a short summary in Sec.IV.

\section{ A brief review of the LHT model}

The LHT is a nonlinear $\sigma$ model with a global symmetry under
the $SU(5)$ group and a gauged subgroup $[SU(2) \otimes U(1)]^2$.
The $SU(5)$ global symmetry is broken down to $SO(5)$ by the vacuum
expectation value (VEV) of the $\sigma$ field, $\Sigma_0$, given by
\begin{eqnarray}
\Sigma_0=\langle\Sigma\rangle
\begin{pmatrix}
{\bf 0}_{2\times2} & 0 & {\bf 1}_{2\times2} \\
                         0 & 1 &0 \\
                         {\bf 1}_{2\times2} & 0 & {\bf 0}_{2\times 2}
\end{pmatrix}.
\end{eqnarray}
After the global symmetry is broken, there arise 14 Goldstone bosons
which are described by the ``pion'' matrix $\Pi$. The Goldstone
bosons are then parametrized as
\begin{equation}
\Sigma = e^{i\Pi/f}\ \Sigma_0\ e^{i\Pi^T/f}\equiv e^{2i\Pi/f}\
\Sigma_0, \label{sigmaA}
\end{equation}
where $f$ is the breaking energy scale.

The $\sigma$ field kinetic Lagrangian is given by
\cite{ArkaniHamed:2002qy}
\begin{equation}
\mathcal{L}_{\rm K}= \frac{f^{2}}{8} {\rm Tr} | D_{\mu} \Sigma |^2,
\label{kinlag}
\end{equation}
with the $[SU(2)\otimes U(1)]^2$ covariant derivative defined by
\begin{equation}
D_{\mu} \Sigma = \partial_{\mu} \Sigma - i \sum_{j=1}^2 \left[ g_{j}
W_{j\,\mu}^{a} (Q_{j}^{a}\Sigma + \Sigma Q_{j}^{a\,T}) + g'_{j}
B_{j\,\mu} (Y_{j} \Sigma+\Sigma Y_{j}^{T}) \right],
\end{equation}
where $W_{j}^\mu = \sum_{a=1}^{3} W_{j}^{\mu \, a} Q_{j}^{a}$ and
$B_{j}^\mu = B_{j}^{\mu} Y_{j}$ are the heavy $SU(2)$ and $U(1)$
gauge bosons, with $Q_j^a$ and $Y_j$ the gauge generators, $g_j$ and
$g'_j$ are the respective gauge couplings. In the gauge boson
sector, T-parity is introduced as an exchange symmetry between the
gauge bosons of the two different copies of the SM gauge group as
\begin{eqnarray}
 W_{1\mu}^a \longleftrightarrow W_{2\mu}^a,~~~~
 B_{1\mu} \longleftrightarrow B_{2\mu}.
\end{eqnarray}
The light(\emph{L}) and heavy(\emph{H}) gauge fields can be obtained
as
\begin{eqnarray}
 W_L^a &=& \frac{W_1^a + W_2^a}{\sqrt{2}},~~~
 B_L = \frac{B_1 + B_2}{\sqrt{2}}, \nonumber \\
 W_H^a &=& \frac{W_1^a - W_2^a}{\sqrt{2}},~~~
 B_H = \frac{B_1 - B_2}{\sqrt{2}}.
\end{eqnarray}

The electroweak symmetry breaking $SU(2)_L \times U(1)_Y \to
U(1)_{em}$ takes place via the usual Higgs mechanism. The mass
eigenstates of the gauge fields are given by
\begin{eqnarray}
&& W_L^{\pm} = \frac{W_L^1 \mp i W_L^2}{\sqrt{2}},~~~~
  \left( \begin{array}{c} A_L \\ Z_L \end{array} \right)
 = \left( \begin{array}{rc} \cos\theta_W & \sin\theta_W \\
 -\sin\theta_W & \cos\theta_W \end{array} \right)  \left(
 \begin{array}{c} B_L \\ W_L^3 \end{array} \right), \nonumber \\
&& W_H^{\pm} = \frac{W_H^1 \mp i W_H^2}{\sqrt{2}}, ~~~
  \left( \begin{array}{c} A_H \\ Z_H \end{array} \right)
 = \left( \begin{array}{cr} \cos\theta_H & -\sin\theta_H \\
 \sin\theta_H & \cos\theta_H \end{array} \right)
 \left( \begin{array}{c} B_H \\ W_H^3 \end{array} \right),
\end{eqnarray}
where $\theta_{W}$ is the usual Weinberg angle and $\theta_H$ is the
mixing angle defined by
\begin{eqnarray}
\sin{\theta_{H}} \simeq \frac{5 g g^{\prime}}{4(5 g^2 - g^{\prime
2})} \frac{v_{SM}^2}{f^2}.
\end{eqnarray}

At $\mathcal O(v^{2}/f^{2})$ in the expansion of the Lagrangian
(\ref{kinlag}), the mass spectrum of the gauge bosons after EWSB is
given by
\begin {equation}
M_{W_{L}}=\frac{gv}{2}(1-\frac{v^{2}}{12f^{2}}),~~M_{Z_{L}}=\frac{gv}
{2\cos\theta_{W}}(1-\frac{v^{2}}{12f^{2}}),~~M_{A_{L}}=0,
\end {equation}
\begin {equation}
M_{W_{H}}=M_{Z_{H}}=gf(1-\frac{v^{2}}{8f^{2}}),~~M_{A_{H}}=\frac{g'f}{\sqrt{5}}
(1-\frac{5v^{2}}{8f^{2}}),
\end {equation}
where $v=v_{SM}(1+\frac{1}{12}\frac{v_{SM}^2}{f^2})$ and $v_{SM}=
246$ GeV is the SM Higgs VEV.

The global symmetries prevent the appearance of a potential for the
scalar fields at tree level. The gauge and Yukawa interactions that
break the global $SO(5)$ symmetry induce radiatively a
Coleman-Weinberg potential\cite{CW}, $V_{CW}$, whose explicit form
can be obtained after expanding the $\Sigma$ field
\begin{equation}
    V_{CW} = \lambda_{\phi^2} f^2 {\rm Tr}|\phi|^2
    + i \lambda_{H \phi H} f \left( H \phi^\dagger H^T
        - H^* \phi H^\dagger \right)
    - \mu^2 |H|^2
    + \lambda_{H^4} |H|^4,
\end{equation}
where $\lambda_{\phi^2}$, $\lambda_{H \phi H}$ and $\lambda_{H^4}$
depend on the fundamental parameters of the model, whereas $\mu^2$,
which receives logarithmic divergent contributions at one-loop level
and quadratically divergent contributions at the two-loop level, is
treated as a free parameter.

The $HZZ$, $HWW$ and $HHZZ$, $HHWW$ couplings involved in our
calculations are modified at $\mathcal O(v^{2}/f^{2})$, which are
given by
\begin{eqnarray}
&&V_{HZ_{\mu}Z_{\nu}}=\frac{e^2v}{6\cos^2\theta_{W}\sin^2\theta_{W}}(3-\frac{v^2}{f^2})g_{\mu\nu},\\
&&V_{HW_{\mu}W_{\nu}}=\frac{e^2v}{6\sin^2\theta_{W}}(3-\frac{v^2}{f^2})g_{\mu\nu},\\
&&V_{HHZ_{\mu}Z_{\nu}}=\frac{e^2}{2\cos^2\theta_{W}\sin^2\theta_{W}}(1-\frac{v^2}{f^2})g_{\mu\nu},\\
&&V_{HHW_{\mu}W_{\nu}}=\frac{e^2}{2\sin^2\theta_{W}}(1-\frac{v^2}{f^2})g_{\mu\nu}.
\end{eqnarray}

\section{Calculation and Numerical results}

In our numerical calculations, the SM parameters are taken as
follows\cite{parameters}
\begin{eqnarray}
\nonumber &&G_{F}=1.16637\times 10^{-5}{\rm GeV}^{-2},
~\sin^{2}\theta_{W}=0.231,~\alpha_{e}=1/128,~m_{H}=125{\rm GeV},\\
\nonumber &&~~~M_{Z}=91.1876{\rm GeV},~m_{b}=4.65{\rm
GeV},~m_{e}=0.51{\rm MeV},~m_{\mu}=105.66{\rm MeV}.
\end{eqnarray}

According to the constraints in Refs.\cite{constraints}, we require
the scale to vary in the range $500$ GeV$\leq f\leq 1500$ GeV.

\begin{figure}[htbp]
\scalebox{0.53}{\epsfig{file=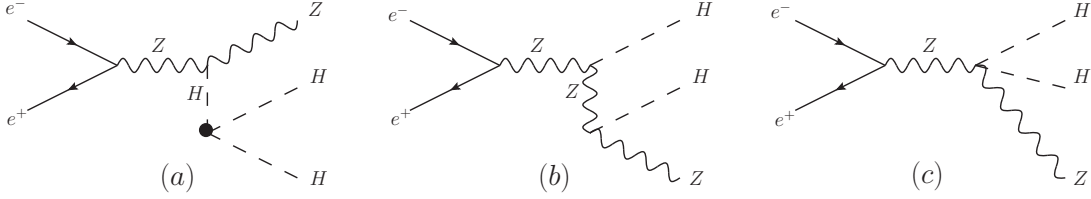}}\vspace{-0.5cm}\caption{
Feynman diagrams for $e^{+}e^{-}\rightarrow ZHH$ at the tree
level.}\label{eezhh}
\end{figure}

\begin{figure}[htbp]
\scalebox{0.45}{\epsfig{file=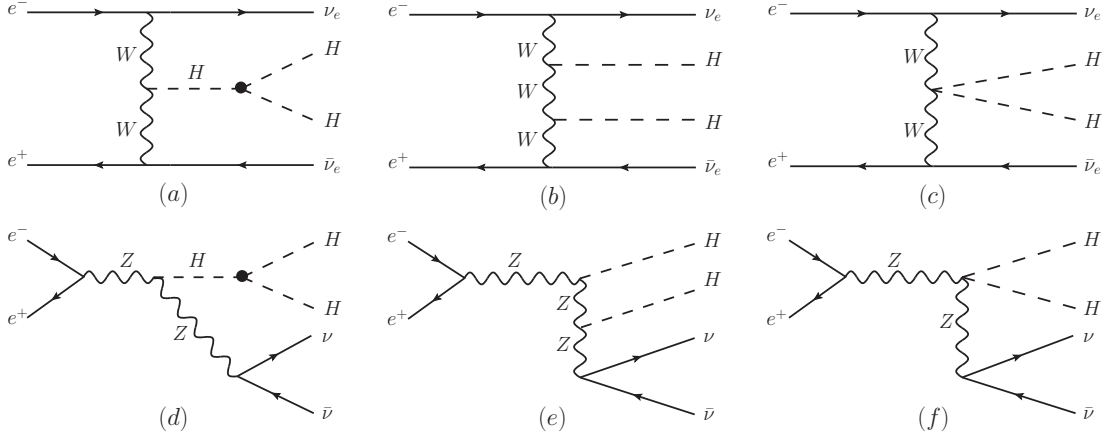}}\vspace{-0.5cm}\caption{
Feynman diagrams for $e^{+}e^{-}\rightarrow \nu\bar{\nu}HH$ at the
tree level.}\label{eevvhh}
\end{figure}

At the tree level, the Feynman diagrams relevant to the process
$e^{+}e^{-}\rightarrow ZHH$ and the process $e^{+}e^{-}\rightarrow
\nu\bar{\nu}HH(\nu=\nu_{e},\nu_{\mu},\nu_{\tau})$ are showed in
Fig.\ref{eezhh} and Fig.\ref{eevvhh}, respectively. In both
processes, we can see that only the first column of the diagrams,
i.e. Fig.1$(a)$, Fig.2$(a)$ and Fig.2$(d)$, are the signal diagrams
which involve the Higgs trilinear self-coupling vertex $HHH$, other
ones are irreducible background diagrams. For the process
$e^{+}e^{-}\rightarrow \nu\bar{\nu}HH$, the $ZZ$-fusion process is
equally or even more important compared with the $WW$-fusion process
at the lower centre-of-mass energy.

\begin{figure}[htbp]
\begin{center}
\scalebox{0.247}{\epsfig{file=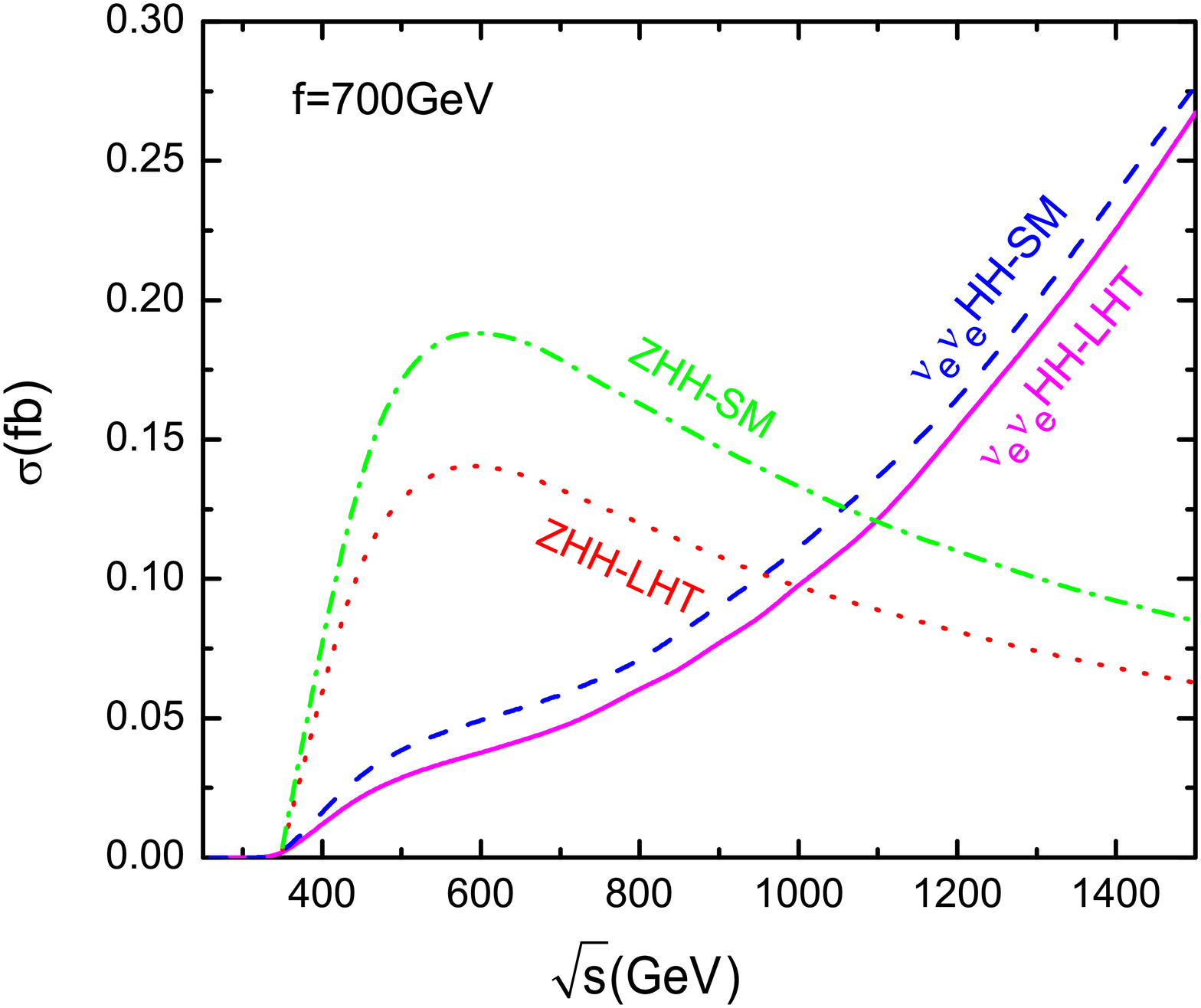}}
\scalebox{0.24}{\epsfig{file=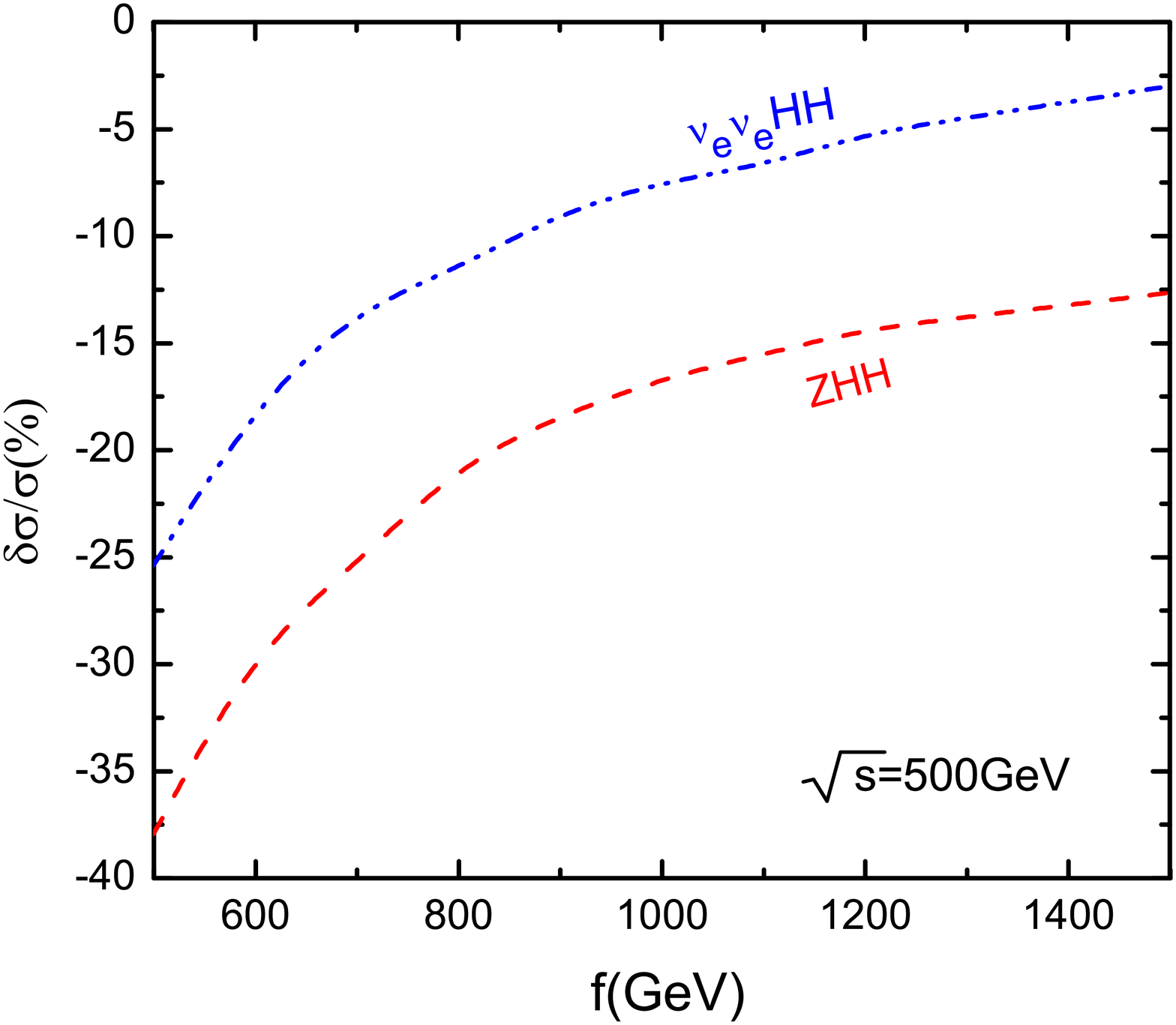}}\vspace{-0.5cm}
\caption{The production cross sections $\sigma$ versus the
center-of-mass energy $\sqrt{s}$ for $f=700$ GeV(left) and the
relative corrections $\delta \sigma/\sigma$ versus the scale $f$ for
$\sqrt{s}=500$ GeV(right).}\label{cross}
\end{center}
\end{figure}
On the left panel of Fig.\ref{cross}, we show the dependance of the
production cross sections $\sigma$ of the processes
$e^{+}e^{-}\rightarrow ZHH$ and $e^{+}e^{-}\rightarrow
\nu\bar{\nu}HH$ on the center-of-mass energy $\sqrt{s}$ for the
scale $f=700$ GeV in the LHT model and the SM, respectively. We can
see that the $e^{+}e^{-}\rightarrow ZHH$ cross section decreases
with increasing center-of-mass energy $\sqrt{s}$ while
$e^{+}e^{-}\rightarrow \nu\bar{\nu}HH$ cross section increases. The
$e^{+}e^{-}\rightarrow ZHH$ cross section has the peak value around
$\sqrt{s}\sim 500$ GeV. For $\sqrt{s}\sim 1$ TeV, the two cross
sections are of the same order of magnitude, with
$e^{+}e^{-}\rightarrow \nu\bar{\nu}HH$ being the larger source of
Higgs boson pairs for $\sqrt{s}\geq 1$ TeV. Since the
$\nu\bar{\nu}HH$ production is peaked in the forward region, it is
important to ensure that an efficient tagging of the $HH \to
b\bar{b}b\bar{b}, W^{+}W^{-}W^{+}W^{-}$ decay can be achieved.

On the right panel of Fig.\ref{cross}, we show the dependance of the
relative corrections $\delta \sigma/\sigma$ of the processes
$e^{+}e^{-}\rightarrow ZHH$ and $e^{+}e^{-}\rightarrow
\nu\bar{\nu}HH$ on the scale $f$ for the center-of-mass energy
$\sqrt{s}=500$ GeV. We can see that the relative correction $\delta
\sigma/\sigma$ of this two processes are both negative and decouple
at the high scale $f$. Considering the lower bound on the scale $f$
from the global fit of the latest experimental data\cite{global
fit}, the relative correction $\delta \sigma/\sigma$ of the process
$e^{+}e^{-}\rightarrow ZHH$ can reach $-30\%\sim -25\%$ and the
relative correction $\delta \sigma/\sigma$ of the process
$e^{+}e^{-}\rightarrow \nu\bar{\nu}HH$ can reach $-16\%\sim -12\%$
for the scale $f$ in the range $600\rm GeV \sim 700\rm GeV$. These
relative correction of the cross section are significant so that
they may be observed at the future $e^{+}e^{-}$ colliders with high
integrated luminosity.

In the following calculations, we will study the process
$e^{+}e^{-}\rightarrow ZHH$ through the
$(l\bar{l})(b\bar{b})(b\bar{b})$ mode and process
$e^{+}e^{-}\rightarrow \nu\bar{\nu}HH$ through
$\nu\bar{\nu}(b\bar{b})(b\bar{b})$ mode. We generate the
parton-level signal and background events with
\textsf{MadGraph5}\cite{mad5}.

\subsection{$e^{+}e^{-}\rightarrow
ZHH\rightarrow (l\bar{l})(b\bar{b})(b\bar{b})$}

\begin{figure}[htbp]
\begin{center}
\scalebox{0.27}{\epsfig{file=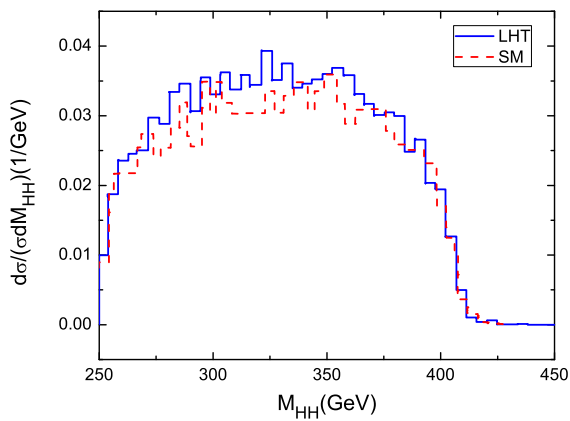}}\vspace{-0.5cm}
\caption{Normalized $M_{HH}$ distributions in the SM and the LHT
through the production of $e^{+}e^{-}\rightarrow ZHH\rightarrow
l\bar{l}b\bar{b}b\bar{b}$ for $\sqrt{s}=500$ GeV, $f=700$
GeV.}\label{zhhll}
\end{center}
\end{figure}

For light Higgs boson masses, the Higgs boson decays predominantly
in a $b\bar{b}$ pair. The $ZHH\rightarrow q\bar{q}b\bar{b}b\bar{b}$
final state benefits from a high statistics with $\sim 35\%$ of the
final states but requires a more complicated analysis. By contrast,
though $ZHH\rightarrow l\bar{l}b\bar{b}b\bar{b}(l=e,\mu)$ represents
only $\sim 5\%$ of the total final state, this topology produces an
easy signature. Therefore, we choose the $l\bar{l}b\bar{b}b\bar{b}$
final state and display some normalized distributions in the LHT
model. The experimental signature is very clean, namely four
$b$-jets (two pairs with invariant Higgs mass) plus $l\bar{l}$ with
invariant $Z$ mass.

In Fig.\ref{zhhll} we display the invariant mass of four $b$-jets
$M_{HH}$ in the SM and LHT model. The $M_{HH}$ distribution is known
to be sensitive to the Higgs boson self-coupling, in particular for
small values of the Higgs-pair mass. Since it is impossible to know
which $b$-jet has to be paired with which $\bar{b}$-jet when
reconstructing the Higgs bosons in the event, here we give the four
$b$-jets invariant mass distribution $M_{HH}$.

The background events mainly come from $e^{+}e^{-}\rightarrow ZZZ
\rightarrow (l\bar{l})(b\bar{b})(b\bar{b})$ and
$e^{+}e^{-}\rightarrow ZZH \rightarrow
(l\bar{l})(b\bar{b})(b\bar{b})$.

\begin{figure}[htbp]
\begin{center}
\scalebox{0.25}{\epsfig{file=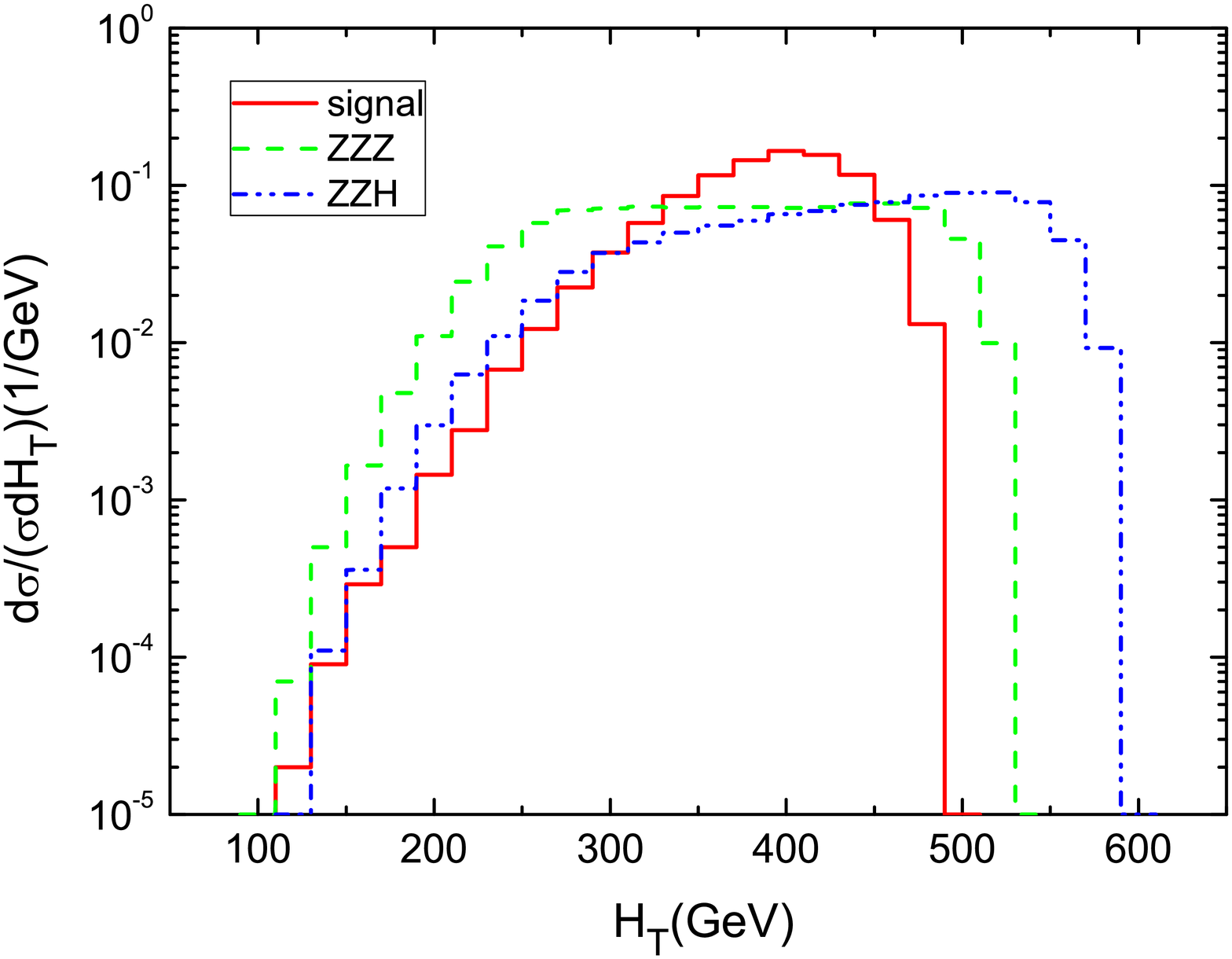}}
\scalebox{0.25}{\epsfig{file=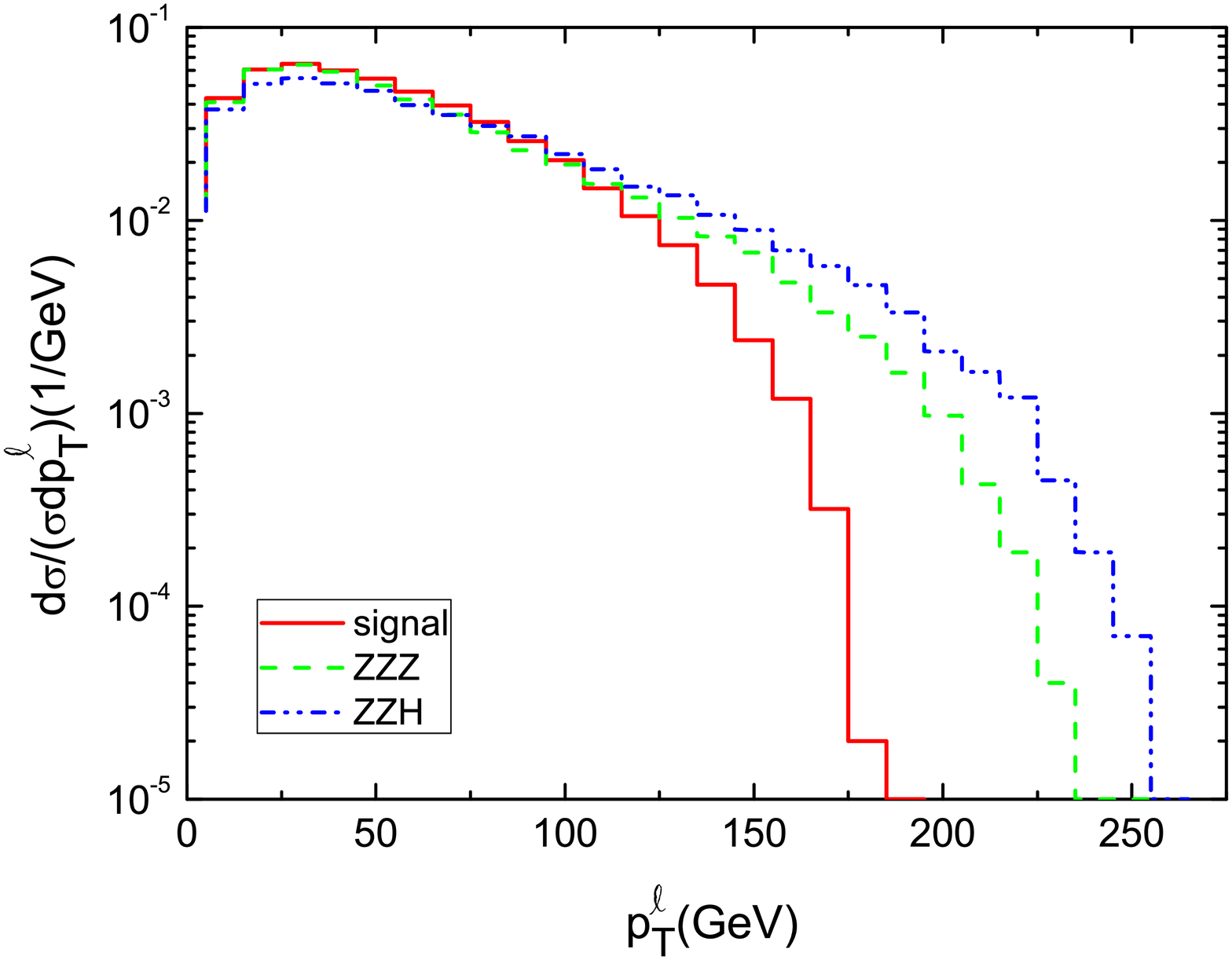}}
\vspace{-0.5cm}\caption{Normalized distributions of $H_{T}$ and
$p_{T}^{l}$ in the signal with $f=700$ GeV and backgrounds for
$\sqrt{s}=500$ GeV.}\label{zhhllsb}
\end{center}
\end{figure}

In Fig.\ref{zhhllsb}, we display the total transverse energy $H_{T}$
and the transverse momentum $p_{T}^{l}$ distributions of
$(l\bar{l})(b\bar{b})(b\bar{b})$ in the signal with $f=700$ GeV and
backgrounds for $\sqrt{s}=500$ GeV. According to Fig.\ref{zhhllsb},
we can impose the cut $H_{T} < 450$ GeV to suppress the backgrounds.
However, due to such the same parton level final states as the
signal, we need more complicated technique and more careful analysis
to distinguish the signal and the backgrounds.

\subsection{$e^{+}e^{-}\rightarrow
\nu\bar{\nu}HH\rightarrow \nu\bar{\nu}(b\bar{b})(b\bar{b})$}

\begin{figure}[htbp]
\begin{center}
\scalebox{0.27}{\epsfig{file=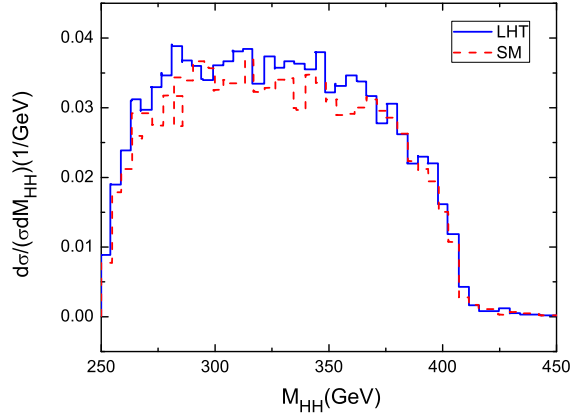}}\vspace{-0.5cm}
\caption{Normalized $M_{HH}$ distributions in the SM and the LHT
through the production of $e^{+}e^{-}\rightarrow
\nu\bar{\nu}HH\rightarrow \nu\bar{\nu}b\bar{b}b\bar{b}$ for
$\sqrt{s}=500$ GeV, $f=700$ GeV.}\label{vvhh}
\end{center}
\end{figure}

Due to the dominant decay mode of Higgs is $H\to b\bar{b}$, the
experimental signature for $e^{+}e^{-}\rightarrow \nu\bar{\nu}HH$ is
then four $b$-jets (two pairs with invariant Higgs mass) plus
missing energy and momentum. The dominant background $\nu\nu bbbb$
mainly comes from $e^{+}e^{-}\rightarrow ZZZ$ and $ZZH$. Likewise,
we display the invariant mass distribution $M_{HH}$ of the four
$b$-jets in Fig.\ref{vvhh}.
\begin{figure}[htbp]
\begin{center}
\scalebox{0.25}{\epsfig{file=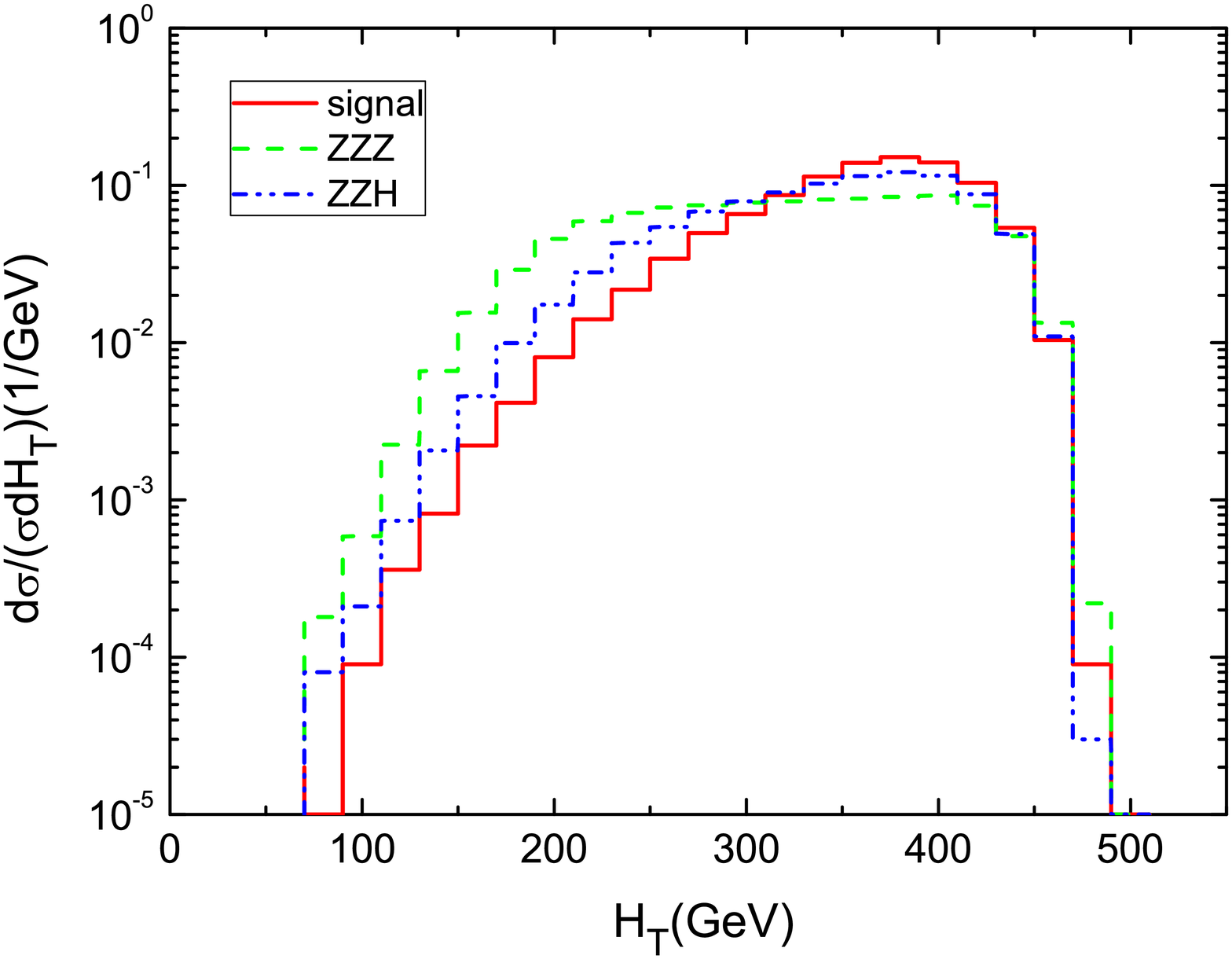}}
\scalebox{0.26}{\epsfig{file=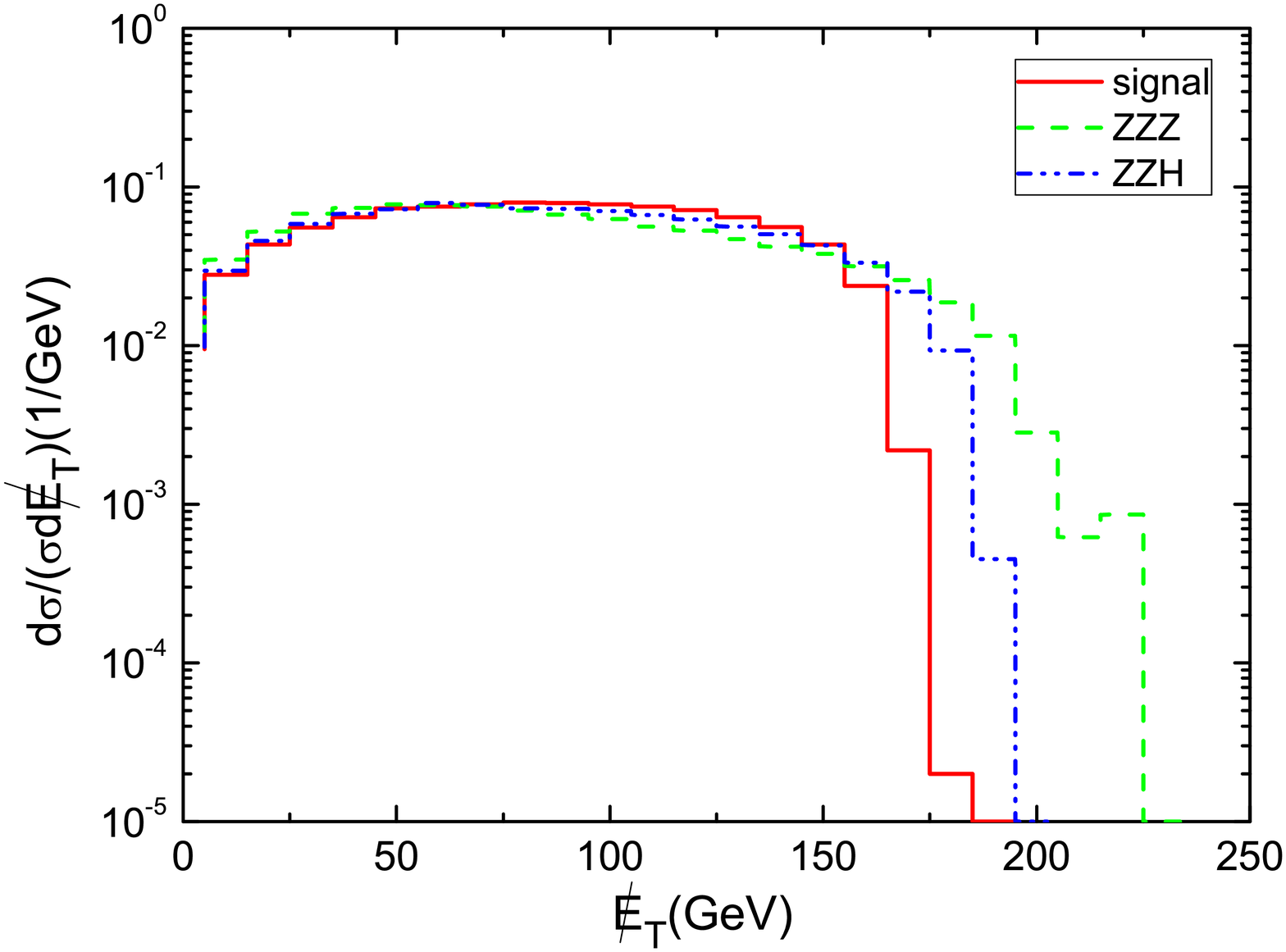}}\vspace{-0.5cm}
\caption{Normalized distributions of $H_{T}$ and $\met$ in the
signal with $f=700$ GeV and backgrounds for $\sqrt{s}=500$
GeV.}\label{vvhhsb}
\end{center}
\end{figure}

In Fig.\ref{vvhhsb}, we display the total transverse energy $H_{T}$
and the missing energy $\met$ distributions of
$\nu\bar{\nu}b\bar{b}b\bar{b}$ in the signal with $f=700$ GeV and
backgrounds for $\sqrt{s}=500$ GeV. According to Fig.\ref{vvhhsb},
we can impose the cut $H_{T} > 300$ GeV to suppress the backgrounds.

\section{summary}

\noindent

In this paper, we studied the double Higgs boson productions at high
energy $e^{+}e^{-}$ colliders in the LHT model. The two main
production channels $e^{+}e^{-}\rightarrow ZHH$ and
$e^{+}e^{-}\rightarrow \nu\bar{\nu}HH$ have been investigated. For
$\sqrt{s}=500$ GeV, we calculated the production cross section and
found that the relative correction of the process
$e^{+}e^{-}\rightarrow ZHH$ can reach $-30\%$ and the relative
correction $\delta \sigma/\sigma$ of the process
$e^{+}e^{-}\rightarrow \nu\bar{\nu}HH$ can reach $-16\%$ when the
scale $f$ is chosen as low as $600$ GeV. This result may be a probe
of the LHT model at the future high energy $e^{+}e^{-}$ colliders.
In order to investigate the observability, the decay modes
$e^{+}e^{-}\rightarrow ZHH \rightarrow l\bar{l} b\bar{b}b\bar{b}$
and $e^{+}e^{-}\rightarrow \nu\bar{\nu}HH\rightarrow
\nu\bar{\nu}b\bar{b}b\bar{b}$ were studied and some distributions of
the signal and background were presented. Due to there is only
slight difference between the signals and backgrounds, more
complicated technique and more careful analysis are needed to
distinguish them.

\section*{Acknowledgement}
This work is supported by the National Natural Science Foundation of
China under grant Nos. 11405047, 11305049 and 11347140, by and the
China Postdoctoral Science Foundation under Grant No.2014M561987 and
the Joint Funds of the National Natural Science Foundation of China
(U1404113).

\end{document}